\documentclass[fleqn,usenatbib]{mnras}

\usepackage{comment}
\usepackage{mathptmx}

\usepackage[T1]{fontenc}

\DeclareRobustCommand{\VAN}[3]{#2}
\let\VANthebibliography\thebibliography
\def\thebibliography{\DeclareRobustCommand{\VAN}[3]{##3}\VANthebibliography}

\usepackage{graphicx}	
\usepackage{amsmath}	
\usepackage{threeparttable}
\usepackage{enumerate}
\usepackage[version=3]{mhchem}
\usepackage{float}
\usepackage{placeins}
\usepackage{ulem}

\newcommand{\cmsq}{\mbox{cm$^{-2}$}}
\newcommand{\cmcub}{\mbox{cm$^{-3}$}}
\newcommand{\NHtwo}{\mbox{$N$(H$_2$)}}
\newcommand{\NOH}{\mbox{$N$(OH)}}

\newcommand{\NCH}{\mbox{$N$(CH)}}

\newcommand{\OHp}{\mbox{OH$^+$}}
\newcommand{\HI}{H {\sc i}}

\title[OH-H$_2$ relation in diffuse clouds]{Investigating the OH-H$_2$ relation in diffuse Galactic clouds}

\author[Rawlins \& Mookerjea]{
Katherine Rawlins$^{1}$\thanks{E-mail: katherine.rawlins@gmail.com}
and Bhaswati Mookerjea$^{2}$\thanks{E-mail: bhaswati@tifr.res.in}
\\
$^{1}$Department of Physics, St. Xavier's College, 5, Mahapalika Marg, Mumbai 400001, India\\
$^{2}$Department of Astronomy \& Astrophysics, Tata Institute of Fundamental Research, Homi Bhabha Road, Mumbai 400005, India\\
}

\date{Accepted XXX. Received YYY; in original form ZZZ}

\pubyear{2023}

\begin{document}
\label{firstpage}
\pagerange{\pageref{firstpage}--\pageref{lastpage}}
\maketitle

\begin{abstract}
We investigate the correlation between OH and H$_2$ column densities in diffuse Galactic clouds, in order to identify potential molecular tracers of interstellar H$_2$. For this, we analyse near-UV spectra extracted from the ESO/VLT archives towards seventeen sightlines (five of them new) with known $N$(H$_2$), along with nine sightlines with no H$_2$ information. \NOH\, shows only marginal correlation with \NHtwo\ (10$^{20}$ to 2 $\times$ 10$^{21}$ cm$^{-2}$), at the 95 per cent confidence level. We use orthogonal distance regression analysis to obtain \NOH\,/\NHtwo\, = (1.32$\pm$0.15) $\times$ 10$^{-7}$, which is $\sim$ 33 per cent higher than the previous estimates based on near-UV data. We also obtain \NCH\,/\NHtwo\, = (3.83$\pm$0.23) $\times$ 10$^{-8}$ and a significant correlation between \NOH\, and \NCH, with \NOH\, = (2.61$\pm$0.19) $\times$ \NCH, both of which are consistent with previous results. Comparison with predictions of numerical models indicate that OH absorption arises from diffuse gas (n$_{\rm H}$ $\sim$ 50 cm$^{-3}$) illuminated by radiation fields $\sim$ 0.5--5 G$_0$, while CH is associated with higher density of 500 cm$^{-3}$. We posit that the apparent dichotomy in the properties of the diffuse clouds giving rise to OH and CH absorption could be due to either (a) the presence of multiple spectroscopically unresolved clouds along the line-of-sight, or, (b) density gradients along the line-of-sight within a single cloud.
\end{abstract}




\section{Introduction}
\label{sec:intro}

Study of the physical and chemical properties of molecular clouds, their formation and evolution is crucial for understanding the star formation history and evolution of a galaxy such as our own, the Milky Way. The first critical step in this whole process is the formation of molecular clouds from diffuse atomic gas. A major hindrance in accessing this step
lies in the fact that the molecular gas is primarily in the form of H$_2$ which has no allowed transitions that are excited in the cold environments of most of the interstellar material. Therefore, CO has long been used as a surrogate tracer of H$_2$ in the Milky Way \citep{Bolatto13, Miville2017}. However, recent observational and theoretical studies suggest that CO is photodissociated to ionized and neutral C atom in a significant fraction of the diffuse molecular gas, with this phase termed as CO-dark molecular gas \citep{Wolfire2010, pineda2013, pineda2014}. In the Milky Way, CO-dark H$_2$ gas is understood to comprise $\sim$ 30 per cent of the molecular mass \citep{pineda2013}. Further, CO is known to get depleted onto dust grains in cold and dense environments \citep{Whittet10}, and chemical fractionation may also influence the abundance of the $^{12}$CO molecule relative to its isotopologues \citep{Sheffer07b}. Hence, observations of chemical species tracing diffuse interstellar clouds are vital in order to understand the transition from atomic to molecular phase. It is also necessary to identify alternative molecular tracers which are accessible over a wide range of molecular hydrogen column densities and interstellar conditions.  

As \citet{Liszt2002} pointed out, there are only a few species {\em viz.,} CH, OH, CN and CO, that can be observed in the optical/UV and radio domains, and hence provide an important bridge between two different ways of studying interstellar chemistry in diffuse clouds. Of these, CH was among the earliest molecular species to be detected in the diffuse ISM \citep{Swings37, McKellar40, Adams41}, and has been proven to have a very well-determined abundance relative to H$_2$. It is found to be nearly constant for all lines of sight with $A_{\rm V} < 3$, with a mean $N$(CH)/$N$(H$_2$) = (4.3$\pm$1.9) $\times 10^{-8}$, and \NCH\ $\propto$ \NHtwo$^{1.00\pm0.06}$ for 10$^{19} <$ $N$(H$_2$) $< 10^{21}$\,\cmsq\ \citep{Liszt2002}. The tight correlation between the CH and H$_2$ abundances is understood based on the higher formation rates of CH in reactions involving molecular hydrogen. Based on a fit, \citet{Sheffer08} constrained $N$(CH)/$N$(H$_2$) to a value of 3.5 $\times 10^{-8}$, which is consistent with the results for dark clouds with $N$(H$_2$) < 6 $\times$ 10$^{21}$ cm$^{-2}$ \citep{Mattila86}. 

OH absorption in the Galaxy has also been studied since long, using observations at radio wavelengths \citep{Weinreb63}, as well as through the ultraviolet electronic transitions \citep{Crutcher76, Chaffee77}. Like CH, the OH molecule has also been explored as a surrogate for H$_2$, particularly in diffuse molecular gas without CO, as the ground-state main lines of OH in the radio regime are readily detectable in translucent/diffuse molecular clouds \citep[e.g.,][]{Magnani1990, Barriault2010}. OH is considered to be a precursor molecule for the formation of CO in diffuse regions \citep{BlackDalgarno1977, Barriault2010}. Models of the diffuse ISM also indicate that OH absorbers with column densities in the range 10$^{13}$ -- 10$^{14}$ cm$^{-2}$ trace CO-dark molecular gas \citep{Balashev2021}. However, in order to establish OH as a bona-fide tracer of H$_2$, the OH/H$_2$ abundance ratio, $X_{\rm OH}$ = $N_{\rm OH}$/$N_{\rm H_2}$, must be constrained. This is a challenge due to the limited direct observations of H$_2$ for sightlines where OH is detected. Based on near-UV observations of six sightlines, previous studies estimated the \NOH/\NHtwo\ ratio to be $\sim$ 1.0 $\times$ 10$^{-7}$ \citep{Liszt2002, Weselak09, Weselak2010}. Additionally, a definite correlation between the OH and CH column densities in diffuse/translucent clouds has been deduced based on a larger sample of twenty-four sightlines, which in turn indicates an indirect correlation between OH and H$_2$ \citep{Mookerjea16, Weselak09, Weselak2010}.

Astrochemical models have indicated values of $X_{\rm OH}$ ranging from 10$^{-7}$ \citep{BlackDalgarno1977} and 1.6 $\times$ 10$^{-8}$ -- 2.9 $\times$ 10$^{-7}$ for $A_{\rm V}$ = 0.6--2.1 mag \citep{vanDishoeckBlack1986}. \citet{Weselak09} were unable to draw conclusions about any underlying correlation between observed column densities of OH and H$_2$ on account of sparse UV data (only six observations). But recent radio observations have obtained a typical value of $X_{\rm OH}$ $\sim 10^{-7}$ \citep{Nguyen18, Rugel2018, Tang2021}. Observational studies detecting higher values of OH abundance of a few times 10$^{-7}$ in molecular boundaries, find the value to decrease towards $X_{\rm OH}$ = 10$^{-7}$ at visual extinctions $A_{\rm V}\geq 2.5$\,mag \citep{Xu2016}. Far-infrared absorption studies of the $^2\Pi_{3/2} J=5/2\leftarrow 3/2$ transition of OH at 2514.3\,GHz measure OH abundances between 10$^{-8}$--10$^{-6}$ \citep{Wiesemeyer16,Jacob2022}. However, \NHtwo\, used by these longer wavelength studies, are not direct measurements. Typically, molecular clouds capable of attenuating the UV radiation field imply that the OH abundance can no longer be constant. Depending on the radiation field at $A_{\rm V}\sim 6$\,mag, oxygen-bearing species are depleted onto ice in the grain mantles, thus affecting the OH abundance \citep{Hollenbach2012}.

Some near-UV Galactic sightlines in the literature are reported to host OH$^+$ along with OH. First observed in the interstellar medium only a few years ago at microwave \citep{Wyrowski10}, UV \citep{Krelowski10} and far-infrared \citep{Neufeld2010} wavelengths, several more near-ultraviolet detections of OH$^+$ along Galactic sightlines have been made subsequently \citep{Porras14, Bacalla19, Weselak20}. Recently, \citet{Weselak20} found that the column densities of OH and OH$^+$ are strongly correlated for the available sample of fourteen sightlines.  

OH has been known to exist in a few intervening absorbers outside the Milky Way too \citep[eg.][]{Chengalur99, Kanekar03}. Improved observational capabilities in the radio regime have brought about renewed interest in probing diffuse extragalactic gas for the presence of OH \citep{Gupta18, Gupta21}. Thus, constraining the OH abundance, $X_{\rm OH}$, in the Galaxy will also be useful for studies of molecular chemistry and abundances in other galaxies and high-redshift environments. 

In this paper, we have performed a combined analysis of archival and literature data on near-ultraviolet absorption measurements of vibrionic transitions of OH along twenty-three sightlines (ten of them new) for which measurements of $N$(H$_2$) are available. This data is combined with data available in literature to study the correlation between OH column densities and column densities of H I, H$_2$, CH and OH$^+$. Observed column densities and/or abundances are compared with the calculations of models for diffuse photodissociation regions in order to, (a) gain insight into the impact of density and radiation field on the abundances, and (b) understand the physical properties of the diffuse interstellar clouds hosting these molecules. 

The spectroscopic data analysis and subsequent results are presented in Section \ref{sec:analysis} and Section \ref{sec:coldens} respectively. Details and results of the numerical models are described in Section \ref{sec:models}, followed by a discussion in Section \ref{sec:disc} and a summary in Section \ref{sec:conc}.

\section{Near-UV Spectroscopy}
\label{sec:analysis}

\subsection{Data \& Basic Analysis Methods}
\label{ssec:data}

In order to detect OH in diffuse interstellar clouds, we identified from literature, potential sightlines in the direction of bright Galactic sources in which, either (i) molecules such as CH and/or CN  have been detected \citep{Sheffer08}, and/or (ii) estimates of $N$(H$_2$) are available \citep{Rachford02, Rachford09, Shull21}. For these selected sightlines, we retrieved reduced high-resolution near-ultraviolet spectra from the science archive of the European Southern Observatory (\url{https://archive.eso.org/scienceportal/home}). All spectra had been obtained using the Ultraviolet and Visual Echelle Spectrograph at Very Large Telescope, Chile, at a resolution higher than 40,000.

Multiple spectra for a particular sightline were first co-added using PyRAF routines \citep{STScI12}. The resultant spectrum was then normalized locally for regions around the absorption lines of interest by performing a linear fit. The observed spectrum was then divided by the fitted continuum, and selected transitions of OH, CH and OH$^+$ (Table\,\ref{tab:transitions}) were identified in the normalized spectrum. 

The equivalent widths of the absorption features were determined as follows:
\begin{equation}
   W = \Delta \lambda \sum_{n = 1}^{M} \left(1-\frac{F}{F_{\rm c}}\right) 
	\label{eq:wr}
\end{equation}
where $\Delta \lambda$ is the spectral dispersion, $M$ is the number of pixels associated with the absorption line, $F$ and $F_c$ refer to the actual flux and the continuum level respectively for a particular pixel. \\
 
The column density values were calculated under the optically thin approximation using:
\begin{equation}
    N = 1.13 \times 10^{20}\frac{W}{\lambda_{\rm r}^2 f_{\rm osc}}
	\label{eq:N}
\end{equation}
where $N$ is the column density, $\lambda_{\rm r}$ is the rest wavelength  of the transition and $f_{\rm osc}$ is the oscillator strength. Table\,\ref{tab:transitions} presents the rest wavelengths and oscillator strengths of the probed transitions.
The uncertainty in equivalent width is estimated using the relation:
\begin{equation}
\sigma^2(W) = \frac{h_\lambda}{S/N} \sum_{j=1}^{M} (\frac{F_j}{F_c})^2 + [\frac{\sigma(F_c)}{F_c}(\delta \lambda - W)]^2
	\label{eq:W_er}
\end{equation}
where $h_\lambda$ is the spectral dispersion per pixel, $F_j$ is the flux in the $j$th pixel, and $\delta$ $\lambda$ is the extent of the absorption line \citep[][Appendix]{Chalabaev1983}. The signal-to-noise ratio, $S/N$, is related to the flux, as follows:
\begin{equation}
\sigma(F_j) = \frac{F_j}{S/N}
\end{equation}
The column density, $N$, has an uncertainty which is estimated in the following manner:
\begin{equation}
    \sigma(N) = \sigma(W) \times N/W
\end{equation}
The total column density of OH is determined by adding the column densities derived from the equivalent widths of the 3078 and 3082\,\AA\ transitions. Similarly, the total CH column density is the sum of the individual column densities obtained from the 3137 and 3143\,\AA\ transitions. 

\begin{table*}
\setlength{\tabcolsep}{4pt}
 \centering
  \caption{Transitions probed for each diffuse Galactic sightline
  \label{tab:transitions}}
   \begin{tabular}{@{}llcccl@{}}
  \hline
  Species & Vibrionic band & Rotational line & Wavelength (\AA) & Oscillator strength & Reference \\
  \hline
  OH & A$^2\Sigma^+$-X$^2\Pi_i$ (0,0) & Q$_1$(3/2)+$^Q$P$_{21}$(3/2) & 3078.443 & 0.00105 & \citet{Weselak09} \\
  & & P$_1$(3/2) & 3081.6643 & 0.000648 & \citet{Weselak09} \\
  CH & C$^2\Sigma^+$-X$^2\Pi_i$ (0,0) & R$_2$ & 3137.576 & 0.00210 & \citet{Lien84} \\
  & & Q$_2$(1)+$^Q$R$_{12}$(1) & 3143.150 & 0.00640 & \citet{Lien84} \\
  OH$^+$ & A$^3\Pi$-X$^3\Sigma^-$ (0,0) & $^r$R$_{11}$(0) & 3583.757 & 0.000527 & \citet{Hodges18} \\
  \hline
\end{tabular}
\end{table*}

\subsection{Sample of OH sightlines}
\label{ssec:sample}

Spectra for thirty diffuse Galactic lines-of-sight were extracted from the archival data and analysed. Along ten of these sightlines, OH was detected and seven among them lie within $\sim$ 10 degrees of the Galactic disk. While the sightline towards HD23180 was previously observed \citep{Chaffee77, Weselak09}, we independently analysed the spectrum following the methods used for the remaining nine new sightlines. It is this updated value that we consider in our further analysis and discussion. 

Table \ref{tab:programid} presents details of the OH sightlines. Even though several more diffuse absorbers were identified and analysed based on known CH, CN or H$_2$ absorption, these are not included as part of the new results presented henceforth in this paper. 

Table\,\ref{tab:coldens} presents the measured equivalent widths of the OH absorption lines at 3078 and 3082 \AA, along with the column densities of OH and CH for the ten OH sightlines. H$_2$ column densities for these absorbers reported in literature are also tabulated \citep{Savage77, Rachford02, Cartledge04, Rachford09, Dirks19, Shull21}. For four of the new sightlines, we have also determined the OH$^+$ column densities based on the $\lambda$3583 absorption feature, and using the improved oscillator strength of \citet{Hodges18}. Two of these are new detections -- the sightlines towards HD62542 and HD168076. OH$^+$ has been previously detected in the other two absorbers, that is, along the line-of-sight to HD41117 and HD179406 \citep{Krelowski10, Bacalla19, Weselak20}, and the column densities we estimate are consistent with the values in literature. The column densities for which S/N $< 3$, are marked with upper limits in Table\,\ref{tab:coldens}, and are not treated in any of the numerical estimates hereafter.

Throughout the paper, "This work" refers to the positive OH detections from the current analysis. Additionally, there are thirteen sightlines in literature for which column densities of OH, CH and H$_2$ are available, (twelve of these with $N$(OH) $>$ 3-$\sigma$), and a further eleven sightlines for which column densities of OH and CH are available but with no corresponding $N$(H$_2$) measurements (nine of these with $N$(OH) $>$ 3-$\sigma$). The OH and CH column densities taken from literature are from \citet{Weselak09, Weselak2010, Bhatt15, Mookerjea16}, with further details of the specific sightlines provided in the footnotes of Table \ref{tab:coldens}. 
Thus, by combining  the sightlines from literature with those we have newly analyzed, the total sample size with S/N $>$ 3$\sigma$ is twenty-six (Table\,\ref{tab:coldens}). \OHp\ is also detected in twelve of the twenty-three primary sightlines for which $N$(H$_2$) is available (three of these with $N$(OH$^+$) $>$ 3-$\sigma$). The measured $E(B-V)$ values towards the OH sightlines range between 0.28 to 1.59\,mag, corresponding to a range of 0.9 to 4.9\,mag of visual extinction $A_{\rm V}$ (for an $R_{\rm V}$ of 3.1).

\section{Correlations between observed quantities}
\label{sec:coldens}

\begin{figure*}
\centering
\includegraphics[width=\textwidth, keepaspectratio]{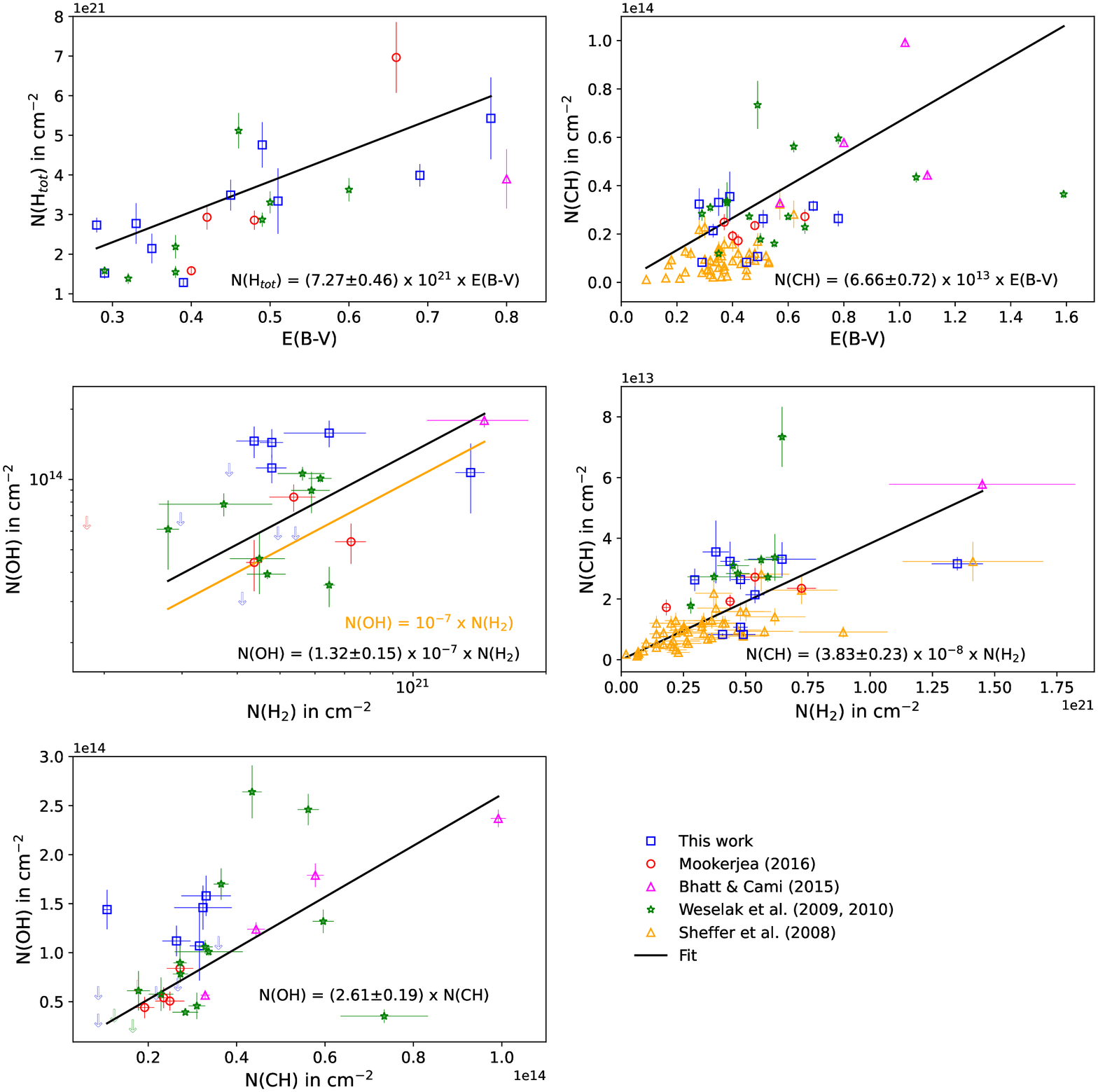}
\caption{Correlations between molecular column densities and $E(B-V)$ along OH and CH lines-of-sight studied in this work and in literature (refer to the common key for all plots). Observational data points in all plots are labelled according to the study which reports OH, except for \citet{Sheffer08}, from which only CH column densities are considered. Additional sightlines from literature have been included for certain correlation plots, as mentioned in the individual panels. The black straight line corresponds to the fit derived by using the method of orthogonal distance regression, which accounts for uncertainties in quantities plotted along both abcissa and ordinate. The equation of the fit is mentioned in the corresponding panel. Table\,\ref{tab:coeffs} presents the calculated Pearson $r$ correlation coefficients.
\label{fig:corrs}}
\end{figure*}

We first explore the correlations between the OH, CH, H$_2$ and H$_{\rm tot}$ column densities and $E(B-V)$ pair-wise (Fig.\ref{fig:corrs}), by calculating the Pearson correlation coefficient. The standard error in the correlation coefficient is estimated using $\sigma_r = (1-r^2)/\sqrt {n-1}$, as done by \citet{Weselak20}. Here, $r$ is the Pearson correlation coefficient and $n$ is the number of data points. While deriving the correlations, wherever possible, we have included data from literature, e.g., data from \citet{Sheffer08} for the $N$(CH)--$N$(H$_2$) relation. 

Table\,\ref{tab:coeffs} presents the correlation coefficients between pairs of column densities and extinction properties, along with the critical values of the Pearson $r$ coefficient that correspond to a 5\% possibility that the correlation factor arises by chance. We find the correlation coefficients for several pairs of quantities to be larger than the critical values (Table \ref{tab:coeffs}).

\begin{table*}
\setlength{\tabcolsep}{4pt}
 \centering
  \caption{OH equivalent widths and column densities of various molecular species for the Galactic sightlines with OH detection
  \label{tab:coldens}}
  \begin{tabular}{@{}lrcccccccccccc@{}}
  \hline
   Source & $E(B-V)$ &  $EW$(3078) & $EW$(3082) & \textit{N}(OH) & \textit{N}(H$_2$) & Ref. & \textit{N}(H I) & Ref. & \textit{N}(CH) & \textit{N}(OH$^+$) & Ref. \\
   & & (m\AA) & (m\AA) & (10$^{13}$ cm$^{-2}$) & (10$^{20}$ cm$^{-2}$) & & (10$^{20}$ cm$^{-2}$) & & (10$^{13}$ cm$^{-2}$) & (10$^{13}$ cm$^{-2}$) & \\
 \hline
 \multicolumn{12}{c}{\bf This work}\\
 \hline
HD23180	& 0.29 &	1.14	$\pm$	0.40	&	1.00	$\pm$	0.23	&	$<$ 3.09	&	4.07	$\pm$	0.61	& 1  &	7.08	$\pm$	0.64	&	11	&	0.83	$\pm$	0.04	&		\ldots	& \ldots  \\	
HD41117	&	0.45 & 2.60	$\pm$	1.38	&	1.62	$\pm$	0.81	&	$<$ 5.89	&	4.90	$\pm$	0.49	&  2  &	25.12	$\pm$	3.77	&  2 &	0.83	$\pm$	0.07	&	5.78	$\pm$	1.01    & \ldots  \\	
HD46202	& 0.49   &	5.40	$\pm$	2.62	&	4.52	$\pm$	2.77	&	14.38	$\pm$	2.01	&	4.79	$\pm$	0.29	&  2  &	38.02	$\pm$	5.70	& 11  &	1.07	$\pm$	0.10	&	\ldots	& \ldots   \\
HD62542	&	0.35 & 6.69	$\pm$	1.19	&	4.43	$\pm$	0.84	&	15.80	$\pm$	2.06	&	6.46	$\pm$	1.36    & 3	&	8.51	$\pm$	2.55	& 3  &	3.31	$\pm$	0.56	&	$<$ 2.56    &  \ldots \\	
HD73882	& 0.69 &	4.35	$\pm$	0.89	&	3.15	$\pm$	0.63	&	10.70	$\pm$	3.54	&	13.49	$\pm$	1.03	&  3 &	12.88	$\pm$	1.93	&  2, 12 &	3.16	$\pm$	0.22	&	\ldots & \ldots    \\	
HD147683 & 0.39	&	4.80	$\pm$	1.17	&	2.97	$\pm$	1.97	&	$<$ 11.00	&	3.80	$\pm$	0.53	& 4  &	5.25    &	13		&	3.55	$\pm$	1.03	&		\ldots	& \ldots	\\	
HD147888 & 0.51	&	2.54	$\pm$	0.96	&	2.14	$\pm$	0.77	&	$<$ 6.76	&	2.95	$\pm$	0.15	& 2  &	27.54	$\pm$	8.26	& 2  &	2.63	$\pm$	0.37	&		\ldots	& \ldots	\\	
HD168076 & 0.78 	&	5.82	$\pm$	1.14	&	2.51	$\pm$	0.47	&	11.22	$\pm$	1.56	&	4.79	$\pm$	0.38	& 3  &	44.67	$\pm$	10.27	& 11  &	2.64	$\pm$	0.32	&	2.29	$\pm$	0.72    &\ldots	\\	
HD179406 & 0.33	&	3.12	$\pm$	0.99	&	1.25	$\pm$	0.78	&	$<$ 5.89	&	5.37	$\pm$	0.38	& 2  &	16.98	$\pm$	5.09	& 2, 14  &	2.14	$\pm$	0.28	&	$<$ 2.59    &\ldots	\\	
HD203532 & 0.28	&	5.07	$\pm$	0.91	&	4.80	$\pm$	1.09	&	14.57	$\pm$	2.25	&	4.37	$\pm$	0.39	& 5  &		18.62	$\pm$	1.68	&   5	&	3.24	$\pm$	0.65	&		\ldots	&	\ldots \\	
\hline
\multicolumn{12}{c}{\bf OH and CH sightlines from literature$^{a,b}$: H$_2$  available}\\

\hline
HD24398	& 0.29 & 	1.67	$\pm$	0.08	&	1.11	$\pm$	0.05	&	3.93	$\pm$	0.18	&	4.68	$\pm$	0.47	& 6, 7  &		6.46	$\pm$	0.52	& 11	&	2.84	$\pm$	0.29	&	$<$ 0.50    & 19	\\	
HD27778	& 0.38 &	5.30	$\pm$	0.15	&	2.20	$\pm$	0.10	&	10.10	$\pm$	0.40	&	6.17	$\pm$	0.37	& 3  &	9.55	$\pm$	2.86    & 3	&	3.37	$\pm$	0.77	&	\ldots	&	\ldots \\
HD34078	& 0.49 &	1.72	$\pm$	0.27	&	0.86	$\pm$	0.21	&	3.53	$\pm$	0.69	&	6.46	$\pm$	0.19	& 7, 8  &		15.85	$\pm$	1.74	&   11	&	7.34	$\pm$	0.99	&\ldots	&  \ldots \\	
HD110432 & 0.40	&	1.81	$\pm$	0.41	&	1.28	$\pm$	0.34	&	4.41	$\pm$	1.09	&	4.37	$\pm$	0.18	& 3  &	7.08	$\pm$	1.06	& 15  &	1.92	$\pm$	0.22	&	$<$ 0.67    & 20, 21	\\	
HD147933 & 0.46	&	3.63	$\pm$	0.27	&	2.02	$\pm$	0.31	&	7.83	$\pm$	0.88	&		3.73	$\pm$	1.07		&  1 &		43.65	$\pm$	3.93	& 11  &	2.73	$\pm$	0.17	&  \ldots &	\ldots		\\	
HD149757 & 0.32	&	2.01	$\pm$	0.67	&	1.25	$\pm$	0.32	&	4.58	$\pm$	1.35	&		4.49	$\pm$	0.64		& 1  &		4.90	$\pm$	0.49	& 11  &	3.10	$\pm$	0.19	&		\ldots & \ldots	\\	
HD151932 & 0.60	&	4.46	$\pm$	0.71	&	2.13	$\pm$	0.54	&	8.97	$\pm$	1.80	&		5.89	$\pm$	0.59		& 9  &		24.55	$\pm$	2.70	& 11  &	2.72	$\pm$	0.15	&	$<$ 1.82    &	20, 21 \\	
HD152236 & 0.66	&	3.50	$\pm$	0.44	&	2.41	$\pm$	0.34	&	8.40	$\pm$	1.12	&	5.37	$\pm$	0.64	& 2  &		58.88	$\pm$	8.83		&  2, 11 &	2.72 $\pm$ 0.30		&	$<$ 2.00	&  20, 21 \\	
HD152249 & 0.42	&	2.81	$\pm$	1.00	&	1.82	$\pm$	0.80	&	$<$ 6.53	&		1.81	$\pm$	0.12		& 10  &		25.70	$\pm$	3.08    & 11	&	1.72 $\pm$ 0.26			&	1.42	$\pm$	0.46	& 20, 21  \\	
HD152270	& 0.50 &	1.95	$\pm$	0.64	&	2.13	$\pm$	0.70	&	6.12	$\pm$	2.01	&		2.79	$\pm$	0.16		& 10 &		27.54	$\pm$	2.75		& 10  &	1.78 $\pm$ 0.26			& \ldots	&  \ldots \\	
HD154368 & 0.80	&	9.25	$\pm$	0.58	&	4.01	$\pm$	0.31	&	17.90	$\pm$	1.20	&	14.45	$\pm$	3.74	& 3  &	10.00	$\pm$	0.50		& 3, 16  &	5.78	$\pm$	0.19	&	$<$ 2.00	&  20, 21 \\	
HD170740 & 0.48	&	1.97	$\pm$	0.43	&	1.73	$\pm$	0.31	&	5.41	$\pm$	1.06	&	7.24	$\pm$	0.58	& 3  &	14.13	$\pm$	2.12		& 11  &	2.35	$\pm$	0.13	&	$<$ 1.33	& 20, 21  \\	
HD210121 & 0.38	&	4.45	$\pm$	0.28	&	3.03	$\pm$	0.23	&	10.60	$\pm$	0.70	&	5.62	$\pm$	0.68	& 3  &	4.27	$\pm$	0.64	& 3, 17  &	3.29	$\pm$	0.17	&	$<$ 1.33	&  20, 21 \\
\hline
\multicolumn{12}{c}{\bf OH and CH sightlines from literature$^{a,b}$: H$_2$ not available}\\
\hline	
BD-145037 & 1.59	&	5.74	$\pm$	0.43	&	5.71	$\pm$	0.60	&	17.00	$\pm$	1.60	&		\ldots		&  \ldots &		\ldots		&  \ldots &	3.65	$\pm$	0.17	&	2.48	&	19	\\	
HD114213 & 1.06	&	9.86	$\pm$	0.96	&	8.30	$\pm$	0.87	&	26.40	$\pm$	2.70	&		\ldots		& \ldots  &		\ldots		& \ldots  &	4.35	$\pm$	0.22	&	\ldots	&	\ldots	\\	
HD147889 & 1.02	&	12.19	$\pm$	0.45	&	5.38	$\pm$	0.20	&	23.70	$\pm$	0.90	&		\ldots		& \ldots  &		50.12   & 12	&	9.92	$\pm$	0.17	&	\ldots	&	\ldots \\	
HD148688 & 0.55	&	0.92	$\pm$	0.45	&	0.81	$\pm$	0.54	&	$<$ 2.53	&		\ldots		& \ldots  &		\ldots		& \ldots  &	1.61	$\pm$	0.17	&	\ldots	&	\ldots \\	
HD154445 & 0.37	&	2.01	$\pm$	0.31	&	1.52	$\pm$	0.33	&	5.07	$\pm$	0.96	&		\ldots		& \ldots  &		10.00	$\pm$	1.00	& 11  &	2.49	$\pm$	0.33	&	$<$ 0.83    & 20, 21	\\	
HD154811 & 0.66	&	2.43	$\pm$	0.70	&	1.64	$\pm$	0.50	&	5.77	$\pm$	1.71	&	\ldots			& \ldots  &		\ldots		& \ldots  &	2.29 $\pm$ 0.28			&	\ldots	&	\ldots	\\	
HD161056 & 0.62	&	10.71	$\pm$	0.61	&	6.78	$\pm$	0.50	&	24.60	$\pm$	1.60	&		\ldots		&  \ldots &		16.98	$\pm$	3.40	& 11  &	5.62	$\pm$	0.24	&	$<$ 3.00    &	20, 21 \\	
HD163800 & 0.57	&	2.00	$\pm$	0.17	&	1.85	$\pm$	0.12	&	5.67	$\pm$	0.41	&		\ldots		&  \ldots &		\ldots		& \ldots  &	3.29	$\pm$	0.10	&	$<$ 2.00	&  20, 21 \\	
HD164794 & 0.36	&	1.55	$\pm$	0.40	&	1.00	$\pm$	0.65	&	$<$ 3.59	&	\ldots			& \ldots  &		19.50	$\pm$	1.36	& 11  &	1.19 $\pm$ 0.25			&	\ldots	&	\ldots	\\	
HD169454 & 1.10	&	6.24	$\pm$	0.25	&	2.90	$\pm$	0.22	&	12.40	$\pm$	0.70	&		\ldots		& \ldots  &		\ldots		& \ldots  &	4.44	$\pm$	0.20	&	\ldots	&	\ldots	\\	
HD172028 & 0.78	&	5.05	$\pm$	0.50	&	4.09	$\pm$	0.35	&	13.20	$\pm$	1.20	&		\ldots		&  \ldots &		17.30	$\pm$	0.80		&  18  &	5.96	$\pm$	0.24	& \ldots  &	\ldots	\\
\hline
\end{tabular}

\flushleft

1 - \citet{Savage77}, 2 - \citet{Rachford09}, 3 - \citet{Rachford02}, 4 - \citet{Dirks19}, 5 - \citet{Cartledge04}, 6 - \citet{Wannier99}, 7 - \cite{Wolfire08}, 8 - \cite{Boisse05}, 9 - \cite{Rice2018}, 10 - \citet{Marggraf04}, 11 - \citet{Diplas94}, 12 - \citet{Fitzpatrick90}, 13 - \citet{Velusamy17}, 14 - \cite{Hanson92}, 15 - \cite{Rachford01}, 16 - \cite{Snow96}, 17 - \cite{Welty92}, 18 - \citet{Adamkovics05}, 19 - \citet{Bacalla19}, 20 - \citet{Krelowski10}, 21 - \citet{Weselak20}

$^a$ OH equivalent widths and column densities for the HD147889, HD154368, HD163800 and HD169454 sightlines are based on \citet{Bhatt15}. For the remaining sightlines, the values are adopted from \citet{Weselak09, Weselak2010}. \\
$^b$ CH column densities for the HD152236, HD152249, HD152270, HD154811 and HD164794 sightlines were obtained using the 3886 and 3890 {\AA} transitions \citep{Weselak09, Weselak2010}. All other CH column density values from literature were determined through the 3137 and 3143 {\AA} transitions. Of these, column densities for HD24398, HD27778, HD34078, HD110432, HD148688 and HD154445 are based on \citet{Mookerjea16}, while the rest are from \citet{Bhatt15}.

\end{table*}

\begin{table}
\setlength{\tabcolsep}{4pt}
 \centering
  \caption{Pearson correlation coefficients between the OH, CH, H$_2$ and H$_{\rm tot}$ column densities and $E{(B-V)}$. Correlations are based on data from the present analysis and from literature \citep{Weselak09, Weselak2010, Bhatt15, Mookerjea16}. Additional data, where included, are mentioned below the table. The critical value corresponds to the $r$-value which confirms existence of correlation with 95\% confidence. 
  \label{tab:coeffs}}
  \begin{tabular}{@{}lccr@{}}
  \hline
  Relation & Pearson coefficient & $n^\dagger$ & Critical $r$ \\
  &$r$ & & for $p=0.05$ \\
  \hline
  $N$(CH) vs $N$(H$_2$) & 0.66$\pm$0.07 & 73$^a$ & 0.23 \\
  $N$(OH) vs $N$(CH) & 0.51$\pm$0.15 & 26$^b$ & 0.39 \\
  $N$(OH) vs $N$(H$_2$) & 0.44$\pm$0.20 & 17$^b$ & 0.48 \\
  $N$(H$_2$) vs $E(B-V)$ & 0.55$\pm$0.15 & 23 & 0.41\\
  $N$(H$_{\rm tot}$) vs $E(B-V)$  & 0.73$\pm$0.10 & 23 & 0.41 \\
  $N$(CH) vs $E(B-V)$  & 0.59$\pm$0.05 & 84$^a$ & 0.22 \\
  $N$(OH) vs $E(B-V)$ & 0.55$\pm$0.14 & 26$^b$ & 0.39 \\
  \hline
\end{tabular}
\flushleft
$^\dagger$ n: Number of data points\\
$^a$ Includes data from \cite{Sheffer08}. \\
$^b$ Only $N$(OH) measurements with S/N $>$ 3-$\sigma$ are included.\\
\end{table} 

\subsection{CH, OH and H$_{\rm tot}$ correlations with $E(B-V)$}
\label{ssec:eb-v}

The twenty-three sightlines with H$_2$ column density measurements show $N$(H$_2$) between (0.8--14.45) $\times$  $10^{20}$\,\cmsq, and $N$(\HI) between (4.27--58.88) $\times$ $10^{20}$\cmsq. Most of the sightlines are predominantly atomic, with only four (toward HD73882, HD152236, HD154368 and HD210121) sightlines having molecular hydrogen fraction exceeding 50\%. The total hydrogen column density, $N$(H$_{\rm tot}$) = \textit{N}(\HI) + 2 \textit{N}(H$_2$), is strongly correlated with $E(B-V)$ with a Pearson $r$ = 0.73, and the fitted relation between the  total hydrogen nuclei column density and $E(B-V)$ is $N$(H$_{\rm tot}$) = (7.27$\pm$0.46) $\times$ 10$^{21}$ \cmsq $E$(B-V). \citet{Bohlin78} had famously obtained the frequently quoted constant of proportionality, $N$(H$_{\rm tot}$) = 5.8 $\times$ 10$^{21}$ $E(B-V)$ for lines-of-sight with $E(B-V)$ values not exceeding 0.4\,mag.  In a more recent work, \citet{Liszt14} has obtained the relation $N$(\ion{H}{i}) = 8.3 $\times$ 10$^{21}$\,\cmsq $E(B-V)$ for primarily atomic lines of sight with $E(B-V)$ $<$ 0.1 mag. The lines of sight analysed here, though primarily atomic, have significantly larger $E(B-V)$ values, which have mostly been measured using UV extinction \citep{Rachford09, Shull21}. The \NCH--$E(B-V)$ scatterplot shows a definite correlation between the two quantities and a negative intercept on the y-axis which is consistent with the results towards dark clouds \citep[][and references therein]{Liszt2002}. We obtain a relation of \NCH\ = (6.66$\pm$0.72) $\times$ $10^{13}$ $\times$ $E(B-V)$ when fitted with a straight line without an intercept. This compares well within uncertainties with \NCH\ = 0.5 $\times$ 3 $\times$ 10$^{-8}$ $\times$ 5.8 $\times$ $10^{21}$\,$E(B-V)$\,\cmsq\ = 8.7 $\times$ 10$^{13}$\,$E(B-V)$\,\cmsq\, which is an estimate of the maximum CH column density for $X_{\rm CH}=3 \times 10^{-8}$ if all gas along the line-of-sight were in the form of H$_2$. OH also has significant moderate correlation with $E(B-V)$, with a Pearson coefficient of 0.55$\pm$0.14.    

\subsection{OH and CH correlations with H$_2$}
\label{ssec:h2}

Based on the analysis of a compilation of \NCH\ and \NHtwo\ data, \citet{Liszt2002} had derived $N{\rm (CH)}$ = (4.3$\pm$1.9) $\times$ 10$^{-8}$ $N{\rm (H_2)}$ for 10$^{20}$ < $N$(H$_2$) < 2 $\times$ 10$^{21}$\,\cmsq. These column densities were collated from the results of UV and optical stellar absorption studies available at the time. Further, \citet{Sheffer08} obtained a value of \NCH\ = 3.5 $\times 10^{-8}$\,\NHtwo\,, also based on ultraviolet spectra. More recently, \citet{Weselak19} reported \NHtwo/\NCH = (2.01$\pm$0.09) $\times$ 10$^7$ with a negative intercept on the y-axis. This yields \NCH\ = (4.98$\pm$0.22) $\times$ 10$^{-8}$ \NHtwo. Using an orthogonal distance regression suitable for the analysis of data with uncertainty in both x- and y-axes, we obtain an equivalent relation with a factor of (3.83$\pm$0.23) $\times$ 10$^{-8}$, which is consistent with the result of \citet{Sheffer08} and \citet{Liszt2002} within the limits of uncertainties. 

Estimates of the OH column densities based on the measurements of the hyperfine transitions of the $^2\Pi_{3/2} J=3/2$ ground state of OH in the 18-cm band \citep{Rugel2018,Tang2021}, as well as using the group of hyperfine transitions $^2\Pi_{3/2} J=5/2\leftarrow 3/2$ at 2514.3\,GHz \citep{Wiesemeyer12,Wiesemeyer16} are available in the literature. {The $N$(H$_2$) values for the OH sightlines in these studies are not direct measurements, but are scaled estimates obtained from $N$(CH) or $N$(HF) \citep{Wiesemeyer12, Wiesemeyer16}, $^{13}$CO \citep{Rugel2018}, or $E$(B--V) \citep{Tang2021}. Thus, we do not include the far-infrared and radio measurements in our analysis of the OH--H$_2$ relation. Analysing the seventeen UV sightlines in our study with OH and H$_2$, we determine a correlation coefficient of 0.44$\pm$0.20, which is less than the critical value of 0.48 for the Pearson r when p = 0.05. Applying orthogonal distance regression, we obtain $N$(OH)/$N$(H$_2$) = (1.32$\pm$0.15) $\times 10^{-7}$, while the median value obtained for the same ratio is 1.56 $\times$ 10$^{-7}$. Both these values are significantly higher than the previous estimate of $\sim$ 1.0 $\times$ 10$^{-7}$ \citep{Liszt2002, Weselak09, Weselak2010}.}

The scatter plot between $N$(OH) and $N$(CH) shows reasonably good correlation, with the exception of three sightlines -- HD34078, HD114213 and HD161056, with $E(B-V)$ = 0.49, 1.06 and 0.62 respectively. The first of these sightlines has very low OH column density, while the other two show significantly higher values of $N$(OH) (Fig\,\ref{fig:corrs}). The median of the observed values of \NOH/\NCH\ is 3.04, and the fitted value of $N$(OH)/$N$(CH) is 2.61$\pm$0.19. These estimates are consistent with results in literature \citep{Liszt2002, Weselak2010, Mookerjea16}. The \NOH/\NCH\, ratio for individual sightlines in our sample ranges from 0.4 to 13.5, with only two out of the twenty-six lines-of-sight having values outside the range 1--5 (sightlines with S/N $>$ 3 only).

\section{Numerical models of the diffuse interstellar clouds}
\label{sec:models}

OH$^+$, the precursor to the formation of OH, can be produced through two different channels \citep{Draine2011, Gerin2016}. One route is through the fast exothermic reaction between O and H$_3^+$ :  ${\rm O + H_3^{+}} \longrightarrow {\rm OH^{+} + H_2}$. The H$_3^+$ necessary for this reaction is generated by cosmic ray ionization of H$_2$, that is, ${\rm H_2 + CR \longrightarrow H_2^+ + e^{-}}$, followed rapidly by ${\rm H_2^+ + H_2 \longrightarrow H_3^+ + H}$. The second channel for OH$^+$ formation occurs through charge exchange between oxygen and hydrogen: ${\rm O + H^{+}} \longrightarrow {\rm O^{+} + H}$, followed by ${\rm O^{+} + H_2} \longrightarrow {\rm OH^{+} + H}$ \citep{Liszt2003}. Once OH$^+$ is available, the reaction ${\rm OH^{+} + H_2}\longrightarrow {\rm H_2O^{+} + H}$ occurs, forming a water ion and a hydrogen atom. The newly created H$_2$O$^+$, in 20\% of the cases, forms OH and H in a dissociative recombination reaction with a free electron: ${\rm H_2O^+ + e^-} \longrightarrow {\rm OH + H}$. The principal way in which  OH is produced in diffuse and translucent clouds, is similar to the previous reaction sequence in that it involves dissociative recombination and starts with H$_2$O$^+$, but takes a few more steps to form OH \citep{Singh1980}: ${\rm H_2O^{+} + H_2} \longrightarrow  {\rm H_3O^{+} + H}$; ${\rm H_3O^{+} + e^{-}} \longrightarrow {\rm OH + H_2}$ (14\%) or ${\rm H_3O^{+} + e^-} \longrightarrow  {\rm OH + H + H}$ (60 \%). The OH thus formed, is primarily destroyed by photodissociation: ${\rm OH + h \nu} \longrightarrow  {\rm O + H}$.

\subsection{Modelling technique}
\label{ssec:approach}
In order to constrain the physical environment of the diffuse Galactic clouds associated with the observed OH and CH absorption features, we set up a grid of numerical models using the spectral synthesis code \textsc{cloudy}, version c17.02 \citep{Ferland17}. Each model assumes a diffuse cloud of uniform density with the plane-parallel geometry of a photodissociation region \citep{Tielens85}, and illuminated from both sides by an ISRF ($\chi$), with intensities expressed in units of the Habing FUV radiation field, $G_0$ \citep{Habing68}. We consider solar metallicity and cosmic ray ionization rate of 2 $\times$ 10$^{-16}$\,s$^{-1}$, estimated as the mean Galactic value by \citet{Indriolo07}. Graphite and silicate dust grains are included in the model calculations, as per the size distribution proposed by \citet{Mathis77}. Metal depletion on grain surface is also considered in accordance with the overall elemental and grain abundances. We have run a grid of models with different sets of values of total density of hydrogen nuclei, n$_{\rm H}$, and intensity of the ISRF, $\chi$. For each set of parameter values, we model six clouds with different H$_2$ column densities that span the entire range of \NHtwo\ seen in our OH sightline sample. The extent of the modelled absorbers thus, depends on the corresponding H$_2$ column density. The {\sc cloudy} calculations stop at these specified values of $N$(H$_2$) (Table \ref{tab:paras}), and the corresponding column densities for all other chemical species are then noted down and used for comparison with the observational data. Table \ref{tab:paras} presents all parameter values considered for the diffuse cloud models.

\subsection{Physical conditions in the diffuse clouds}
\label{ssec:phycon}

\begin{figure*}
\centering
\includegraphics[width=\textwidth, keepaspectratio]{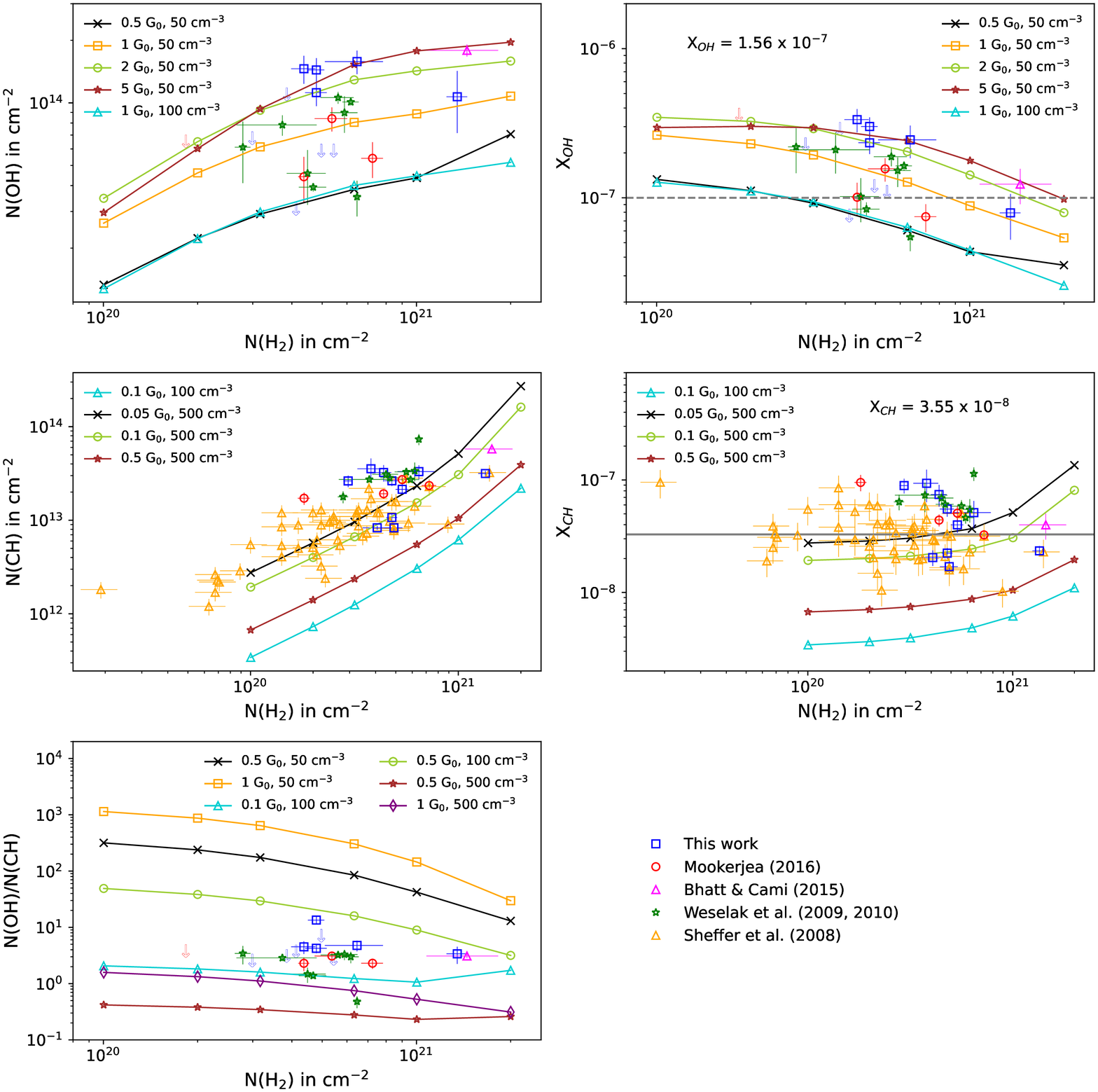}
\caption{Comparison of the observed column density and abundance relations of OH and CH with the corresponding predictions from {\sc cloudy} models. Physical properties of these diffuse Galactic clouds, that is, the intensity of radiation field and hydrogen density, can be constrained from these plots. All models assume solar metallicity and the Galactic background cosmic ray ionization rate of 2 $\times 10^{-16}$\, s$^{-1}$ \citep{Indriolo07}. The X$_{\rm OH}$ and X$_{\rm CH}$ values mentioned in the panels are median values. OH and CH observations are from this work and literature (refer to the common key for all plots). Observational data points in all plots are labelled according to the study which reports OH, except for \citet{Sheffer08}, from which only CH column densities are considered.
\label{fig:models}}
\end{figure*}

The temperature in the coolest, molecular part of these modelled clouds lies in the range 16 to 30 K for the models corresponding to the largest column densities of H$_2$. The extent of the absorbers is less than or up to a few parsecs, and the value of $A_{\rm V}$ in these models is within $\sim$ 2 mag. These values are consistent with the expectations for diffuse/translucent clouds \citep{Snow06}. 

We have compared the observed scatterplots of \NOH, \NCH, \NOH/\NCH, $X_{\rm OH}$ and $X_{\rm CH}$ as functions of \NHtwo\ with the corresponding predictions from the diffuse cloud models explored using {\sc cloudy} (Fig.\,\ref{fig:models}). The grid of  models span the range of \NHtwo\ covered by the OH and CH data presented here and in literature \citep{Weselak09, Weselak2010, Bhatt15, Mookerjea16}. Additional data points from observations available in literature have also been included if they lie in the aforementioned range of \NHtwo. 

The observed $N$(OH)--$N$(H$_2$) correlation can be reproduced by diffuse cloud models with number density 50\,\cmcub\ and FUV radiation field intensity, $\chi$, ranging between 0.5 to 5 G$_0$, with the predicted value of \NOH\ for the same value of \NHtwo\ increasing as the radiation field intensity increases (Fig. \ref{fig:models}). Although we find that the $n_{\rm H}$ = 100\,\cmcub\ models with $\chi = 1$\,G$_0$ predict lower values of $N$(OH), these models could, in principle, reproduce the observed higher values of $N$(OH) for larger values of ISRF intensity. The models with $n_{\rm H}$ = 500\,\cmcub\ however produce significantly lower values of \NOH, and hence, are not plotted. Thus, based on the model predictions, OH absorption appears to arise primarily from extremely diffuse ($n_{\rm H}$ $\sim$ 50\,\cmcub) interstellar clouds. 

In the plot of OH abundance (X$_{\rm OH}$) relative to H$_2$, we have also shown a horizontal line corresponding to the value of X$_{\rm OH}$ = 10$^{-7}$ that is widely used in literature. We find that the X$_{\rm OH}$ values estimated from this work and literature \citep{Weselak09, Weselak2010, Bhatt15, Mookerjea16}, lie in the range 0.5--5 $\times$ 10$^{-7}$, on both sides of this nominal value of 10$^{-7}$. Although not included in the comparison with the {\sc cloudy} models constructed by us, the radio measurements of the 18-cm OH band by \citet{Rugel2018} and the far-infrared measurements of \citet{Wiesemeyer12,Wiesemeyer16} also agree with this range of $X_{\rm OH}$. But the values from \citet{Tang2021} consistently show values well in excess of 10$^{-7}$, going up to 4 $\times$ 10$^{-6}$. However, these FIR and radio estimates of X$_{\rm OH}$ are not based on direct H$_2$ measurements, and are therefore, not combined with the values obtained from our near-UV analysis. 

The lower density models which are consistent with the $X_{\rm OH}$ values observed in the near-UV \citep[This work and][]{Weselak09, Weselak2010, Mookerjea16} show an almost constant $X_{\rm OH}$ with increasing \NHtwo\ up to a column density $\sim 7 \times 10^{20}$\,\cmsq. This is followed by a decreasing trend at still larger values of \NHtwo. The models with $n_{\rm H}$ = 500\,\cmcub\, however, predict $X_{\rm OH}$ significantly lower than the observed values. The value of  $X_{\rm OH}$ predicted by the  model with $n_{\rm H}$ = 50\,\cmcub\ and $\chi$ = 0.5\,G$_0$ is similar to the value predicted by the model with $n_{\rm H}$ = 100\,\cmcub\ and $\chi$ = 1\,G$_0$, thus indicating that $X_{\rm OH}$ depends on $\chi/n_{\rm H}$. 

In case of the \NCH--\NHtwo\ relation, the models predict values consistent with observations only for high density and low ISRF intensity, that is, for models with n$_{\rm H}$ = 500\,\cmcub\ and $\chi$ between 0.05 and 0.1\,G$_0$. Further, \NCH\ is seen to increase with decreasing values of radiation field intensity. For models with $\chi$ = 0.5\,G$_0$ and $n_{\rm H}$ = 100\,\cmcub, the values of predicted \NCH\ are much lower than the corresponding observed values. With the exception of one line-of-sight, the CH absorption features appear to arise from interstellar clouds with $n_{\rm H}$ = 500\,\cmcub, although the corresponding values of observed \NOH\ far exceed the values predicted by these models. 

Similar to the observed variation of \NCH\ with \NHtwo, the observed abundance of CH ($X_{\rm CH}$) is best explained by models with $n_{\rm H} = 500$\,\cmcub. The observed median value of X$_{\rm CH}$ is 3.55 $\times$ 10$^{-8}$, with all the points lying between $\sim$ 10$^{-8}$ and $10^{-7}$. 

We attempt to identify models that consistently explain both observed \NOH\ and \NCH, by comparing the \NOH/\NCH\ ratio as a function of \NHtwo. The results suggest that models with $n_{\rm H}$ = 100\,\cmcub\ and $\chi$ = 0.1--0.5\,G$_0$ reproduce the observed trends. However, on comparing the observed and model-predicted trends of \NOH\ and \NCH\ as functions of \NHtwo\ individually, it appears that absorption due to the two molecules must arise from clouds with densities different by as much as a factor of 10. The models with $n_{\rm H}$ = 50\,\cmcub\ and $n_{\rm H}$ = 500\,\cmcub, that respectively reproduce the observed \NOH\ and \NCH\ values separately, predict higher and lower values of \NOH/\NCH\, respectively. \citet{Mookerjea16} had reproduced the observed correlation between \NOH\ and \NCH\ by considering models with $n_{\rm H}$ = 500\,\cmcub, $\chi$ = 5\,G$_0$ and $\zeta_{\rm CR}$ = 10$^{-14}$\,s$^{-1}$. However, as noted by the author, the observations cannot adequately constrain the three free parameters, viz., $n_{\rm H}$, $\chi$ and $\zeta_{\rm CR}$.

\section{Discussion}
\label{sec:disc}

\begin{figure*}
\centering
\includegraphics[width=\textwidth, keepaspectratio]{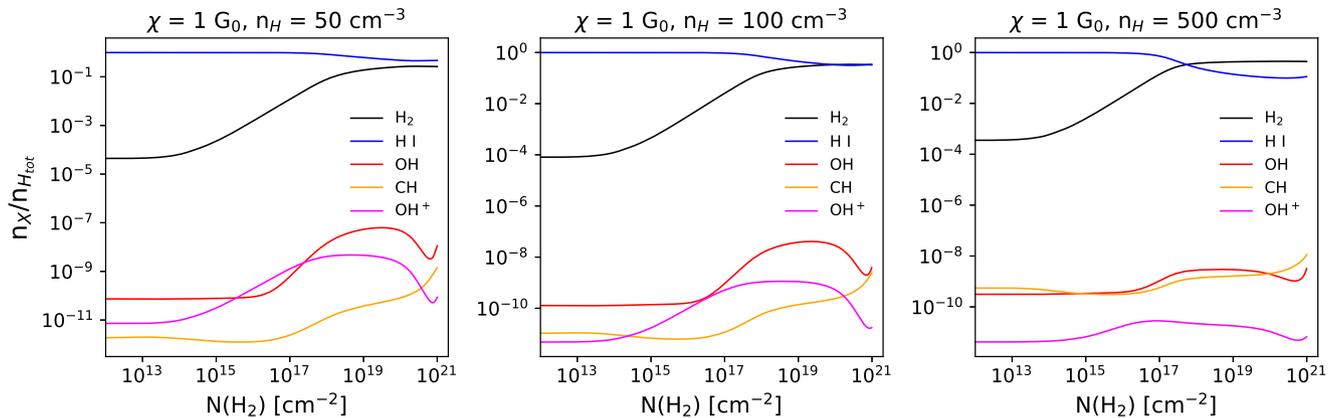}
\caption{The plots indicate the variation in the densities of \ion{H}{i}, H$_2$, CH, OH and OH$^+$ species as a function of N(H$_2$), that is effectively, as a function of depth when one face of the absorber is illuminated. The species density, n$_{\rm X}$, is normalized using the corresponding total hydrogen density, n$_{\rm H_{tot}}$.
\label{fig:abundance}}
\end{figure*}

The observed \NOH\--\NHtwo\ relation can be reproduced by {\sc cloudy}-based FUV-illuminated diffuse cloud models with $n_{\rm H} = 50$\,\cmcub\ and $\chi$ = 1\,G$_0$. This is consistent with the results obtained by \citet{Balashev2021} through their analytical and numerical models, though we note that they consider an ISRF equivalent to the Mathis field \citep{Mathis1983} and apply a lower cosmic ray ionization rate of 3 $\times$ 10$^{-17}$ s$^{-1}$. While \citet{Balashev2021} probe only the OH chemistry in the diffuse ISM, we also investigate CH and OH$^+$ through our models. The observed \NCH\ is reproduced only for molecular gas with densities of $n_{\rm H} = 500$\,\cmcub\ irradiated by much weaker radiation fields, in contrast to the physical conditions deduced from the \NOH\ trends.   

Fig. \ref{fig:abundance} shows various molecular abundance profiles obtained from {\sc cloudy} models when radiation is incident on one side of the absorbing cloud. Using this, we further probe the molecular contributions from different regions within the absorbing cloud by tracing the changes in species abundances from the illuminated face of the cloud to the innermost shielded regions. When radiation is allowed to be incident on both sides of the cloud, similar abundance profiles are expected as a function of depth from each illuminated face of the absorber. For models with $n_{\rm H}$ = 50 and 100\,\cmcub, the OH abundance can be clearly seen as steadily increasing with \NHtwo\ and attaining a peak value of $\sim$ 10$^{-7}$ at \NHtwo\ $\sim$ 5 $\times$ 10$^{19}$\,\cmsq. In comparison, when $n_{\rm H}$ = 500\,\cmcub, the OH abundance has values $<10^{-8}$ for \NHtwo\ < 10$^{21}$\,\cmsq. Along similar lines, the models with lower densities of 50 and 100\,\cmcub\ show CH abundances slowly increasing with \NHtwo, but always being less than 10$^{-9}$ over the entire cloud. Only at the higher density of 500\,\cmcub\ does the abundance increase to values of 10$^{-8}$ or more, for increasing values of \NHtwo. 

We propose two possible reasons for the clear dichotomy in the derived densities of gas giving rise to the observed OH and CH column densities. The first possibility is that the OH and CH absorption arise from multiple clouds overlapping along the line of sight. If the velocities do not differ significantly, it would not be possible to distinctly identify the individual absorption lines at the resolution of the near-UV measurements. The second possibility is that both OH and CH absorption arise from the same spherical cloud with density gradients along the line-of-sight, such that OH is detected primarily in the outer layers of the cloud while the CH absorbing layers are seated deeper within the denser part of the cloud. Layered clouds with the molecular content increasing inward, akin to this latter possibility, were earlier invoked by \citet{Liszt1996} to explain the line profiles of OH and \HI\ along the line of sight. 

Fig. \ref{fig:abundance} also shows the abundance of OH$^+$ predicted by the models. For models with $n_{\rm H}$ = 50\,\cmcub\ and $\chi = 1$\,G$_0$, the abundance is found to increase up to $\sim 10^{-8}$ for depths into the cloud corresponding to \NHtwo\ $\sim$ $10^{19}$ cm$^{-2}$, and then decreases with further increase in depth. For the higher-density models, the abundance of OH$^+$ is much lower than a few times 10$^{-8}$.

The models which reproduce the observed OH and CH column densities do not simultaneously predict consistent OH$^+$ column densities. We infer that it is unlikely that the observed OH$^+$ absorption features arise entirely due to the diffuse molecular clouds that give rise to OH absorption, or the denser clouds in which most of the CH exists. In fact, previous observations have suggested that \OHp\ resides primarily in clouds that are predominantly atomic \citep{Neufeld2010, Indriolo15}, since \OHp\ is rapidly destroyed in reaction with H$_2$. It is possible that the observed OH$^+$ column densities are indicative of high-energy inputs in the form of cosmic ray ionization or shocks. As there are only four OH$^+$ detections with S/N $>$ 3 (two of these new), a detailed investigation of OH$^+$ is beyond the scope of this paper.

\section{Summary}
\label{sec:conc}

We have studied the correlation between OH and H$_2$ column densities using NUV spectroscopic observations, to probe the suitability of OH as a surrogate for H$_2$ in  the diffuse interstellar clouds traced  by these observations. To this effect, we have presented here a new determination of OH column densities for five sightlines along which direct estimates of H$_2$ column densities are available, and have thereby increased the sightlines for which such data is available to seventeen. We have also included in our study, an additional nine sightlines with no reported H$_2$ column densities, in order to explore the correlation between OH, CH and OH$^+$.

We find marginal correlation between the OH and H$_2$ column density measurements at the 95 per cent confidence level, for \NHtwo\ in the range 10$^{20}$ to 2 $\times$ 10$^{21}$ cm$^{-2}$. We determine \NOH/\NHtwo\, = (1.32$\pm$0.15) $\times$ 10$^{-7}$, which is substantially higher than previous estimates of the ratio. The observed $N$(CH) shows significant correlation with the measured \NHtwo, as also with \NOH. We derive \NOH/$N$(CH) = 2.61$\pm$0.19, which is consistent with values found in literature \citep[][and references therein]{Liszt2002}. We conclude that while \NOH\ correlates well with \NCH, the correlation between \NOH\ and \NHtwo\ is not so clear from the near-UV measurements. Thus, to establish OH as a reliable tracer for H$_2$, it is necessary to extend the study to more sightlines with direct measurement of \NHtwo, and also to higher values of \NHtwo.

Based on numerical models of photodissociation regions, we conclude that CH absorption arises from clouds with densities higher than those which produce OH absorption. We propose two possibilities to explain the situation --  either CH and OH absorption arise from separate clouds of different densities lying along the same line-of-sight, or OH absorption arises from less dense portions of clouds whose denser regions are associated with CH absorption. We further infer that the observed OH$^+$ column densities cannot be explained by the physical conditions that give rise to OH and CH, thus suggesting that cosmic rays and/or other high-energy input mechanisms such as shocks may have a possible role in its origin.

\section*{Acknowledgements}
The authors thank the referee for insightful comments which helped improve the manuscript significantly. KR is grateful to TIFR for hospitality during academic visits related to this work. BM was supported by funding from the Department of Atomic Energy, Government of India, under Project Identification No. RTI 4002. This work makes use of spectra observed with the Ultraviolet \& Visual Echelle Spectrograph on the Very Large Telescope at Cerro Paranal, Chile, and made available through the ESO science archive. Software used for data analysis: Python \citep{vanRossum09}, PyRAF \citep{STScI12}, Numpy \citep{Harris20}, Matplotlib \citep{Hunter07}, Scipy \citep{Virtanen20}, Astropy \citep{Price-Whelan22}, Pandas \citep{Pandas22}. 

\section*{Data Availability}

The results of this paper are based on data retrieved from the ESO telescope science archive, and numerical models generated using the code {\sc cloudy}. No new observations have been carried out. Further details about the data analysis and/or numerical models can be made available on request to the authors.



\bibliographystyle{mnras}
\bibliography{oh_refs} 



\appendix
\section{Details of {\textsc uves} archival data and {\textsc cloudy} models}

\FloatBarrier
\begin{table}
\setlength{\tabcolsep}{4pt}
 \centering
  \caption{The diffuse Galactic sightlines in which OH absorption is detected in the present study}
  \label{tab:programid}
  \begin{tabular}{@{}llll@{}}
  \hline
  Source & Program ID & Galactic longitude & Galactic latitude \\
  & & (deg) & (deg) \\
  \hline  
HD 23180  & 194.C-0833$^a$      & 160.3637  & -17.7399\\ 
HD 41117  & 194.C-0833$^a$      & 189.6918  &	-0.8604\\ 
HD 46202  & 096.D-0008$^b$      & 206.3134  & -2.0035 \\ 
HD 62542  & 099.C-0637$^c$      & 255.9152  & -9.2371\\     
HD 73882  & 194.C-0833$^a$      & 260.1816  & 0.6431\\   
HD 147683 & 194.C-0833$^a$      & 344.8565 & 10.0888\\  
HD 147888 & 194.C-0833$^a$      & 353.6469 & 17.7092\\ 
HD 168076 & 194.C-0833$^a$      & 16.9371 & 0.8375\\    
HD 179406 & 194.C-0833$^a$      & 28.2285 & -8.3119\\   
HD 203532 & 194.C-0833$^a$      & 309.4589 & -31.7397 \\ 
\hline
\end{tabular}

$^a$ PI: Cox, Nick, $^b$ PI: Hubrig, S., and $^c$ PI: Nieva, Maria-Fernanda
\end{table}

\begin{table}
\centering
\caption{Values of physical parameters considered in the models of diffuse interstellar clouds set up using {\sc cloudy}
\label{tab:paras}}
\begin{tabular}{@{}ll@{}}
\hline
Physical parameter  &  Values adopted \\
\hline
Intensity of radiation field, $\chi$ & 0.05, 0.1, 0.5, 1, 2, 5, 10, 50, and 100 G$_0$\\
Total hydrogen density, n$_{\rm H}$ & 50, 100, and 500 cm$^{-3}$\\
Metallicity & Z$_{\odot}$ \\
Dust composition \& abundance & ISM graphites and silicates\\
Cosmic ray ionization rate, $\zeta_{\rm CR}$ & 2 $\times$ 10$^{-16}$ s$^{-1}$\\
\textit{N}(H$_2$) & 20, 20.3, 20.5, 20.8, 21, and 21.3 (log cm$^{-2}$) \\
\hline
\end{tabular}
\end{table}


\bsp	
\label{lastpage}
\end{document}